\documentstyle[preprint,prl,eqsecnum,aps,epsf]{revtex}

\begin{document}

\preprint{}

\title{Fusion versus Breakup: Observation of Large Fusion Suppression for 
$^9$Be~+~$^{208}$Pb}
\author{M. Dasgupta$^1$, D.J. Hinde$^1$, R. D. Butt$^1$, R. M. Anjos$^2$, 
A.C. Berriman$^1$,
N.~Carlin$^3$, P.R.S.~Gomes$^2$, C.R. Morton$^1$, J.O. Newton$^1$, 
A. Szanto de Toledo$^3$, and K. Hagino$^4$}
\address{$^1$Department of Nuclear Physics, Research School of Physical 
Sciences and Engineering,\\ Australian National University, Canberra, 
ACT 0200, Australia}
\address{$^2$Instituto de F\'isica, Universidade Federal Fluminense, Av. 
Litoranea, Niter\'oi, RJ, 24210-340, Brazil}
\address{$^3$Instituto de F\'isica, Universidade de Sao Paulo, Caixa Postal 
66318, 05315-970 Sao Paulo,
S.P., Brazil}
\address{$^4$Institute for Nuclear Theory, Department of Physics, University of Washington, Seattle,
WA98915, USA}

\date{\today}

\maketitle

\begin{abstract}
Complete fusion  excitation functions 
for $^{9}$Be + $^{208}$Pb have been measured 
to high precision at near barrier energies. The experimental 
fusion barrier distribution extracted 
from these data allows reliable prediction of the expected complete fusion 
cross-sections. However, the measured cross-sections are 
only 68\% of those predicted. The large cross-sections observed
for incomplete fusion
products support the interpretation that this suppression of fusion is 
caused by $^{9}$Be breaking up into charged fragments 
before reaching the fusion barrier. 
Implications for the fusion of radioactive nuclei are discussed.

\end{abstract}


The recent availability of radioactive beams
has made possible the study of the interactions and 
structure of exotic nuclei far from the line of stability. Unstable  
neutron--rich nuclei having very weakly bound neutrons
exhibit characteristic features such as a 
neutron halo\cite{han95} extending to large radii, associated low--lying 
dipole modes, and a low energy threshold for breakup.
These features may dramatically affect fusion and other 
reaction processes. For fusion to occur, the system must overcome the 
barrier  resulting from the sum of the repulsive Coulomb potential
and the attractive nuclear potential.
Experiments with stable beams have 
shown, however,
that fusion near the barrier is strongly 
affected\cite{bec,das} by intrinsic 
degrees of freedom (such as rotation, vibration)
of the interacting nuclei, whose coupling 
with the relative motion effectively 
causes a splitting in energy of the single, uncoupled fusion barrier. 
This gives rise to a distribution of barrier heights\cite{das82}, 
some higher and some lower in energy than the uncoupled barrier, 
and is manifested most obviously
as an enhancement of the fusion cross--sections at energies near and 
below the average barrier.

In the case of halo nuclei, it is well accepted that 
the extended nuclear matter distribution 
will lead to a lowering of the average fusion barrier, and
thus to an enhancement in fusion cross--sections\cite{tak91} 
over those for tightly bound nuclei. 
The effect of couplings to channels which act as  doorways to breakup
is, however,
controversial\cite{hus1,das1,tak1,sig98}.
Any coupling will enhance the sub--barrier cross-sections, whereas 
breakup may result in capture of only a part of the projectile,
thus suppressing complete fusion.
Model 
predictions\cite{hus1,das1,tak1} however differ in the relative magnitudes 
of enhancement and 
suppression. 
To investigate the effect of the loosely bound neutrons, 
fusion excitation functions in the barrier
region were measured for $^{9,11}$Be~+~$^{238}$U~\cite{BeU} and
$^{9,10,11}$Be~+~$^{209}$Bi~\cite{Yoshida95},  
each study including the reaction with the stable $^{9}$Be for comparison.
The fusion cross-sections 
for $^{10,11}$Be~+~$^{209}$Bi at energies near and below the 
barrier were found to be similar to those for $^{9}$Be, while
 above the barrier the $^{9}$Be induced reaction gave the lowest fusion yield.
It is not obvious whether this is due to 
differing enhancement or suppression 
for the stable and 
unstable projectiles. 

To investigate the effect on fusion of 
couplings
specific to unstable neutron-rich nuclei, 
it is necessary to reliably predict
the cross--sections expected in their absence. Thus, in the above 
cases, definitive conclusions are difficult unless 
fusion with  $^{9}$Be is well understood.
This requires knowledge of the energy of the average fusion 
barrier, and ideally a measurement of the 
distribution of fusion barriers to obtain information on the couplings. 
All this information can be 
obtained from precisely measured fusion cross-sections 
$\sigma_{\mbox{\scriptsize fus}}$, 
by taking the second derivative of the quantity 
E$\sigma_{\mbox{\scriptsize fus}}$ with respect to 
energy E\cite{row91}. This 
function, within certain limits\cite{das,row91}, represents
 the distribution of barrier probability with energy. The 
shape of the experimental barrier distribution 
is indicative of the couplings present and its 
centroid gives the average barrier
position. This information places severe constraints\cite{lei95} on the  
theoretical models.
For this reason, precise measurements, permitting extraction of 
barrier distributions, have resulted in a 
quantitative and self--consistent description
of the fusion cross--sections and barrier distributions 
for a wide range of reactions in which couplings to single--\cite{mor94}, 
double--phonon\cite{ste95}
and rotational states\cite{wei91,van97} are present. 

This Letter reports on precisely measured complete and incomplete fusion 
cross--sections for the reaction of
$^{9}$Be~+~$^{208}$Pb, and utilises the  barrier distribution for 
complete fusion
extracted from these data  
to determine quantitatively the suppression of fusion due to breakup 
of $^{9}$Be.

The experiments were performed 
with pulsed $^{9}$Be beams (1ns on, 1$\mu$s off) in the energy range  
35.0~--~51.0~MeV, from the 14UD tandem accelerator at the
Australian National University. 
Targets were of $^{208}$PbS ($>$99\% enrichment), 
340 -- 400 $\mu$g.cm$^{-2}$ in thickness, evaporated onto 
15~$\mu$g.cm$^{-2}$ C foils.
For normalisation, two monitor
detectors, placed at angles of 22.5$^{\circ}$ above and
 below the beam axis, measured the elastically scattered beam particles. 
Recoiling heavy 
reaction products were stopped in aluminium catcher foils of thickness
360 $\mu$g.cm$^{-2}$, placed immediately behind the target; the mean range 
of the fusion evaporation residues at the maximum bombarding energy is 
130~$\mu$g.cm$^{-2}$. The reaction products 
were identified by their distinctive $\alpha$--energies and 
half--lives~(270~ns to 138~days).
Alpha particles from
short--lived activity (half--life $T_{1/2} \le$~26~sec) were measured in-situ 
during the 1$\mu$s periods between the beam bursts,
using an annular silicon surface barrier detector 
placed 8~cm from the target, at a mean angle of 174$^{\circ}$ to the beam 
direction. These 
were measured at all beam energies using the same target.
An un--irradiated target and catcher 
was used at each energy for determining the cross--section of 
long--lived products ($T_{1/2} \geq$ 24 min).
Alpha particles  from these products 
 were measured using a silicon surface barrier 
detector situated below the annular counter, such that the target 
and catcher could be placed 0.8~cm from the 
detector after the irradiation. The relative solid angles of the two 
detectors were 
determined using the $T_{1/2} =~$24~minute $^{212}$Rn activity.

Fission following fusion was measured during the irradiations using  
two position sensitive multi--wire proportional counters (MWPCs), each with 
active area 
28.4$\times$35.7 cm$^{2}$,
centred at 45$^{0}$ and $-$135$^{0}$ to the beam direction, and 
located 18.0 cm from the target. 
Absolute cross--sections for evaporation residues and fission 
were determined by performing calibrations
at sub-barrier energies in which elastically-scattered 
projectiles were detected in the two 
monitor detectors, the annular 
detector and the 
backward--angle MWPC.

The compound nucleus $^{217}$Rn formed following complete fusion
of $^{9}$Be with $^{208}$Pb cools dominantly by neutron evaporation;
the measured cross--sections for 2n, 3n, 4n and 5n evaporation residues
are shown in Fig.~\ref{fig1}(a). No proton 
evaporation residues were observed.
In addition to the $\alpha$--particles from the decay of  
Rn nuclei, $\alpha$--particles from the decay of
Po nuclei, which are formed as daughters of 
the Rn nuclei following their $\alpha$ decay, were also observed. The 
yields were, however,  much greater than expected from the Rn 
yields, indicating that there is also
a direct population mechanism. Correcting for the 
Rn daughter yields, the cross-sections for the 
direct production of $^{210,211,212}$Po nuclei 
are shown in Fig.~\ref{fig1}(b). In principle the Po nuclei could originate 
from complete fusion followed by $\alpha xn$ evaporation. However the 
shapes of the excitation functions for these
nuclei are distinctly different from those in 
Fig.~\ref{fig1}(a), and are not typical of fusion--evaporation.
For $^{9}$Be~+~$^{208}$Pb, prompt $\alpha$--particles, measured 
in coincidence with $\gamma$--ray transitions in Po nuclei~\cite{drac}, 
showed angular distributions inconsistent with fusion--evaporation, and
production by an incomplete fusion mechanism was inferred~\cite{drac}.
To investigate the origin of the Po yield 
by the $\alpha$--decay technique, the same 
compound nucleus $^{217}$Rn was formed at similar excitation energies 
in the reaction $^{13}$C~+~$^{204}$Hg.
The $\alpha$ spectra were measured between 1.1 and 1.3 times the average 
barrier 
energy. The $^{211,212}$Po $\alpha$--decays, to which the measurement was 
most sensitive, had cross-sections of $<$5~mb, compared with a total of 
$\sim$160~mb for the $^{9}$Be~+~$^{208}$Pb reaction.
Furthermore, the 
fusion cross-sections determined from the sum of the 
$xn$ evaporation 
and fission cross-sections agreed with the predictions of a 
coupled channels calculation
and the Bass model\cite{bass}, indicating that the $xn$ evaporation yield 
essentially exhausts the total evaporation residue cross-section.
Combined, all these observations
show that the direct Po production observed in the 
$^{9}$Be reaction cannot be due to complete fusion.
It is attributed to incomplete fusion, and will be discussed later.
The observed fission cross-sections  were attributed to 
complete fusion of $^{9}$Be~+~$^{208}$Pb, since fission following incomplete fusion should 
be negligible due to the lower angular momentum and 
excitation energy brought in, and the  higher fission barriers
of the resulting compound nuclei. 

Defining complete fusion experimentally as the capture of all the charge of 
the $^9$Be projectile, the complete fusion cross-section at each energy 
was  obtained by 
summing the Rn $xn$ evaporation residue cross-sections and the fission 
cross-section.
The excitation function for complete fusion is shown by the filled circles in 
Fig.~\ref{fig2}(a), whilst Fig.~\ref{fig2}(b) shows 
the experimental barrier distribution 
$d^{2}(E \sigma_{\mbox{\scriptsize fus}} )/dE^{2}$,
evaluated from these data using a point difference formula\cite{lei95} 
with a c.m.~energy step of 1.92 MeV.
The average barrier position 
 obtained from the experimental barrier distribution is 38.3$\pm$0.6 MeV. 
The uncertainty was determined by randomly scattering the measured 
cross-sections, with Gaussian distributions of standard deviation equal to 
those of the experimental uncertainties, and re-determining the centroid.
By repeating this process many times, a frequency distribution for the 
centroid position was obtained, allowing determination of the variance, 
and thus the uncertainty.

To predict the fusion cross-sections expected from the measured barrier 
distribution, realistic  
coupled channels calculations\cite{hag97} were performed using a
 Woods-Saxon form for the nuclear  
potential with a diffuseness 0.63~fm, depth $-$76 MeV and radius parameter
adjusted such that the average barrier energy of these calculations matched 
that measured.
Couplings to the 
5/2$^-$ and 7/2$^-$ states of the K$^\pi$= 3/2$^-$ ground--state 
rotational band\cite{rotbe}
 in $^{9}$Be, and to the 3$^-$, 5$^-$ and 
the double octupole--phonon\cite{double,das97} 
states in $^{208}$Pb were included.
The coupling strengths were obtained from 
experimental data\cite{rotbe,pbbe2}, except 
for  the double octupole--phonon 
states 
in $^{208}$Pb, which were calculated in the harmonic limit.

 The results of these 
calculations 
are shown in Fig.~\ref{fig2}(a) and (b) by the dashed lines. 
They reproduce satisfactorily the asymmetric shape of the measured barrier 
distribution, but the area under the calculated distribution is much 
greater than that measured. 
 The disagreement is necessarily reflected in the cross--sections as well, 
where the calculated values are considerably larger than those measured.
In contrast, for fusion with tightly bound 
projectiles\cite{lei95,mor94,ste95,wei91,van97},
calculations which correctly reproduce the 
average barrier position and the shape of the 
barrier distribution give an extremely good fit to the cross-sections,
as expected. The disagreement for $^{9}$Be~+~$^{208}$Pb,
even though the barrier energies are correctly reproduced, 
suggests the presence of a mechanism hindering fusion. 
Agreement can be achieved only
if the calculated fusion cross-sections are scaled by 
0.68, resulting in 
the full lines in Fig.~\ref{fig2}(a) and (b). This scaling factor 
will be 
model dependent at the lowest energies, as the calculations are 
sensitive to the types of coupling and their strength. 
However, at energies around and above the average barrier, the calculation 
and hence the scaling factor is more robust,
since changes in couplings or 
potential, within the constraints of the measured barrier 
distribution, do not change the 
suppression factor significantly.
The suppression factor of 0.68 has an uncertainty of $\pm$0.07 
arising from the uncertainty in the mean barrier energy. At above barrier 
energies, there is no evidence, within experimental uncertainty, for an energy 
dependence of the suppression factor, but 
a weak dependence cannot be excluded.

The observed suppression of fusion may be related to 
the large yields of $^{212,211,210}$Po which, as shown 
above, do not result from complete fusion.
They can be formed through 
the breakup of $^{9}$Be, probably into
$^{4,5}$He or two $\alpha$ particles and a neutron, with subsequent
absorption of one of the charged fragments by the
$^{208}$Pb. 
The capture of all fragments after breakup cannot be distinguished 
experimentally from fusion without breakup, 
and is included in the complete fusion yield.
Incomplete fusion products following 
the breakup of $^{9}$Be giving $^{6,7,8}$Li were not observed; they are 
unfavoured due to large negative $Q$ values.
The large cross--sections for incomplete fusion, approximately half 
of those for complete fusion,  demonstrate that $^{9}$Be has a 
substantial probability of breaking up into charged fragments.
The sum of the complete and incomplete fusion 
cross-sections is indicated by the hollow circles in Fig.~\ref{fig2}(a). 
They match the predictions of the 
coupled channels fusion calculation, suggesting a direct relationship
between the flux lost from fusion and the incomplete fusion yields.
However, such a simple 
direct comparison is not strictly possible, since
the cross--sections for incomplete fusion may include contributions 
from higher partial waves which may not have led to complete fusion.

The suppression of fusion observed in this experiment is attributed to 
a reduction of flux at 
the fusion barrier radius due to  breakup of the $^{9}$Be projectiles.
Depending on whether the breakup is dominated by the long range Coulomb 
or the short range nuclear interaction, different distributions of 
partial waves for complete fusion should result. Experimental investigations 
of the partial wave distributions are in progress. Comparison of the 
present results 
with those for lighter targets may give additional insights. 
Measurements~\cite{Eck,Brazil} have shown fusion suppression for such 
reactions at energies well above the fusion barrier, although contrary 
results also exist\cite{muk97}. Further measurements for lighter targets 
would be valuable.

In studies of breakup effects for neutron--rich unstable nuclei, the 
focus has been on the neutron separation energy~\cite{sign97}, 
which for $^{11}$Be is 0.50 MeV, compared with 1.67 MeV for $^{9}$Be. This 
led to the expectation that $^{11}$Be induced 
fusion cross--sections would be suppressed compared with those induced by
$^{9}$Be. However this was not borne out by measurement\cite{sig98}.
The present experiment demonstrates that breakup into 
charged fragments affects fusion very significantly.
The two most favourable charged fragmentation channels 
for $^{9,11}$Be are~:
\begin{eqnarray*}
^9\mbox{Be} &\rightarrow &  n + 2\alpha;~  \mbox{Q = --1.57 MeV}\\
    &            & \alpha + ^{5}\mbox{He};~  \mbox{Q = --2.47~MeV}\\
^{11}\mbox{Be} & \rightarrow &  \alpha + ^{6}\mbox{He} + n;~  \mbox{Q = 
--7.91 
MeV}\\
    &            & \alpha  + \alpha + 3n;~  \mbox{Q = --8.89~MeV},
\end{eqnarray*}
 making $^{9}$Be more unstable in this regard than 
$^{11}$Be. Reactions 
with $^{9}$Be 
therefore offer an excellent opportunity to study breakup 
and its effect on fusion, but 
they should not be taken as a stable standard against which to judge the 
breakup effects of their radioactive cousins.

In summary, the precisely measured fusion excitation function for 
$^9$Be~+~$^{208}$Pb, allowing determination of 
the fusion barrier distribution, shows conclusively that complete fusion of 
$^9$Be is suppressed
compared with the fusion of more tightly bound nuclei.
The calculated fusion cross-sections  need to be scaled by a factor 
0.68$\pm$0.07 in 
order 
to obtain a consistent representation of the measured fusion excitation 
function 
and barrier distribution. The loss of flux at the fusion barrier implied by 
this result can be related to the observed large cross--sections for 
Po nuclei, demonstrating that 
$^9$Be has a large probability of breaking up into two helium nuclei, which
would suppress the complete fusion yield. 
These measurements, in conjunction with breakup cross-sections and elastic 
scattering data, should encourage  a complete theoretical description
of fusion and breakup.
Paradoxically,
breakup of the stable $^{9}$Be appears to be more significant than
breakup of the unstable $^{10,11}$Be in influencing the 
fusion product yields.
This conclusion is favourable for using fusion with radioactive 
beams at near--barrier energies to form new, neutron-rich nuclei.

One of the authors~(M.D.) acknowledges the support of a Queen Elizabeth II 
Fellowship. The work of K.H. was
supported by the Japan Society for the Promotion of
Science for Young Scientists; R.M.A., N.C., P.R.S.G. and A.S. de T. acknowledge 
partial support from CNPq, Brazil.

\newpage
\begin{figure}[t]
\caption{(a) The measured cross--sections for 
fission and the production of Rn isotopes, and (b) of Po isotopes 
following the reaction $^9$Be~+~$^{208}$Pb. The dashed 
lines guide the eye.}
\label{fig1}
\end{figure}

\begin{figure}[t,b]
\caption{(a) The 
excitation function for complete fusion (filled circles) and 
(b) the barrier distribution for the reaction
\protect$^{9}$Be+$^{208}$Pb.  
The dashed line is the result of a coupled channels 
calculation (see text) which ignores 
breakup effects. The full line is the same 
calculation scaled by 
0.68. The sum of measured complete and incomplete fusion 
cross-sections is given by the hollow circles.} 
\label{fig2}
\end{figure}

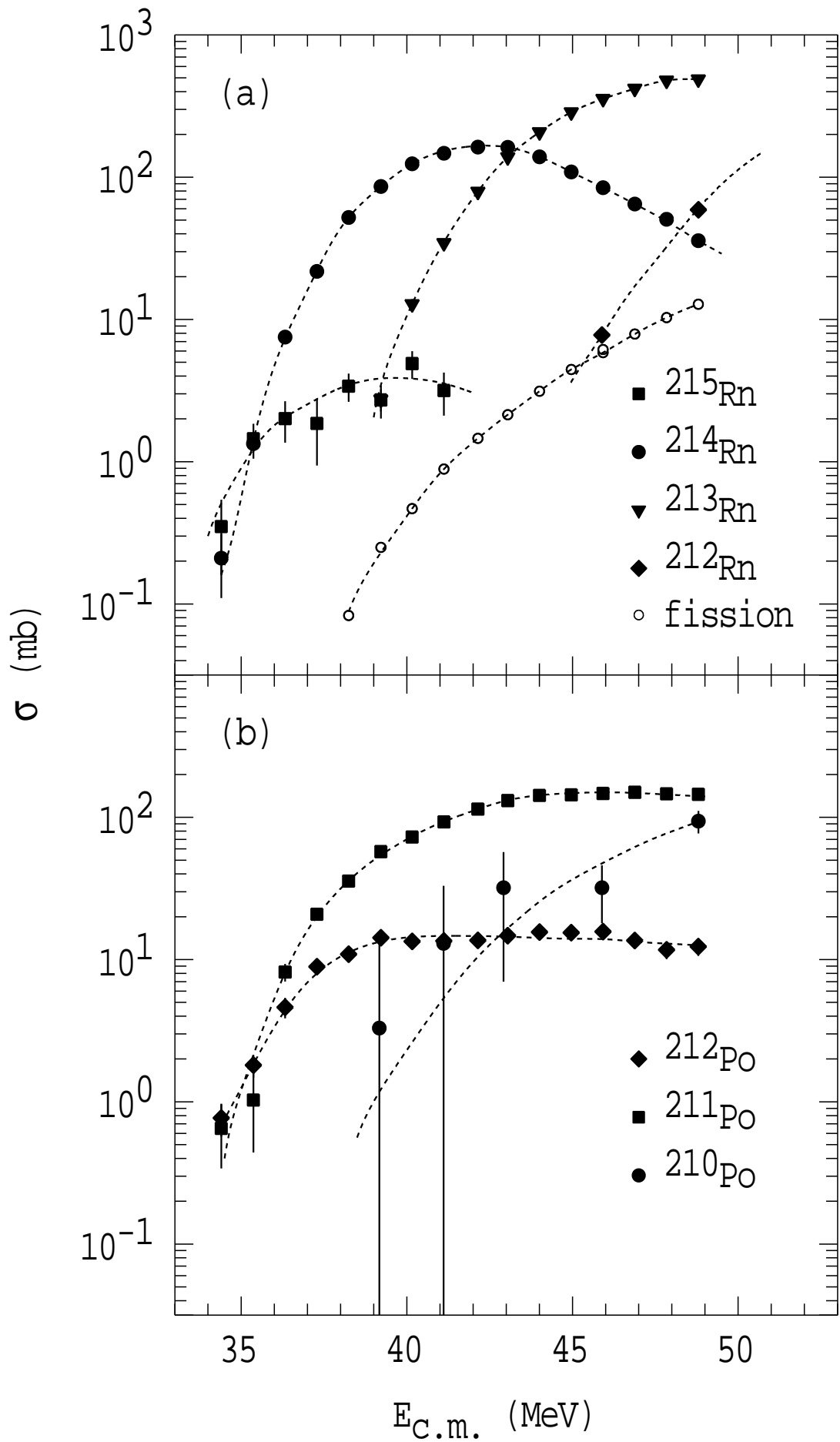
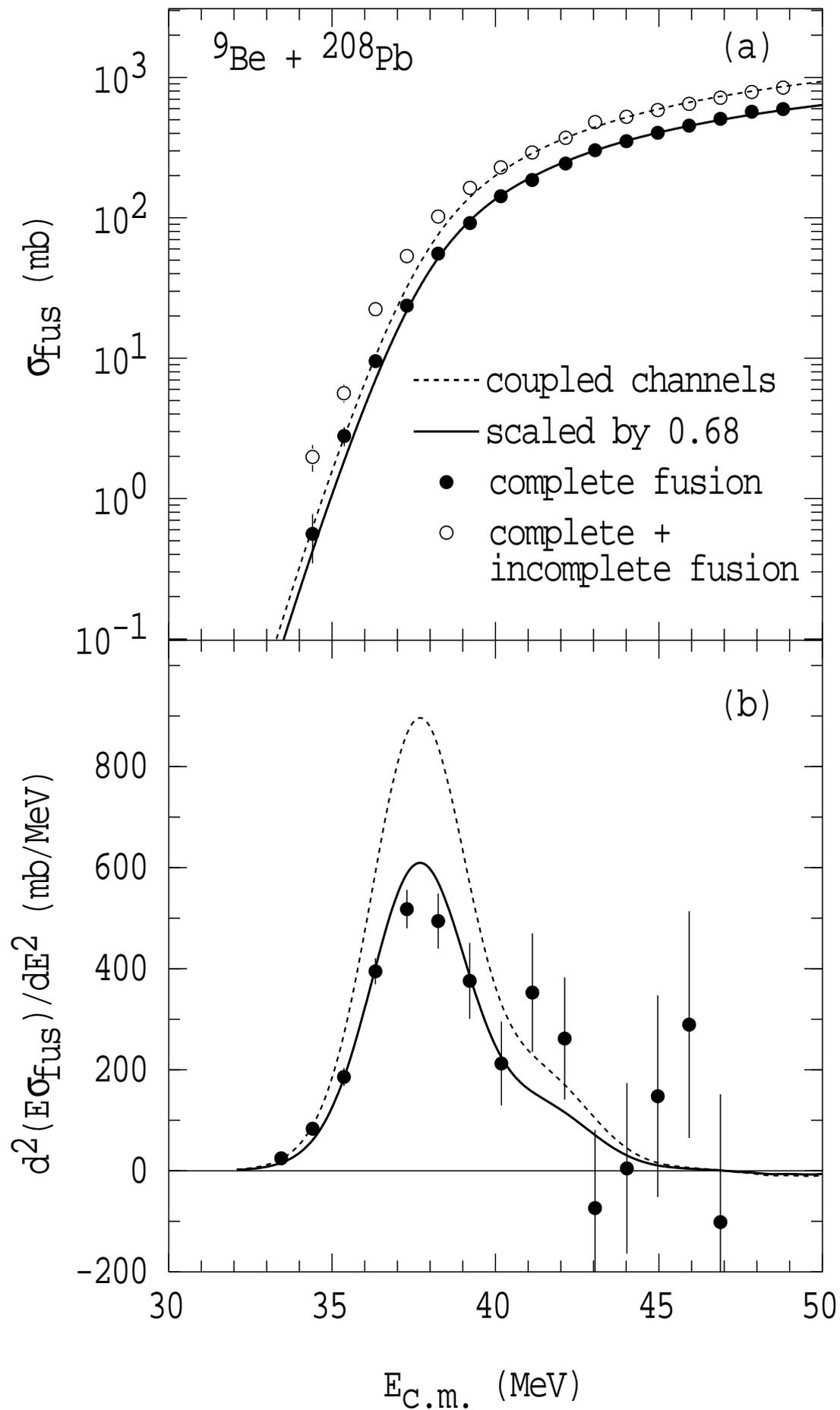
\end{document}